\documentclass[journal=jacsat,manuscript=article]{achemso}
\usepackage{chemformula} 
\usepackage[T1]{fontenc} 
\usepackage{color,soul}

\usepackage{setspace}

\author{David I. Indolese}
\affiliation{Department of Physics, University of Basel, Klingelbergstrasse 82, CH-4056 Basel, Switzerland}
\email{david.indolese@unibas.ch}
\author{Paritosh Karnatak}
\affiliation{Department of Physics, University of Basel, Klingelbergstrasse 82, CH-4056 Basel, Switzerland}
\author{Artem Kononov}
\affiliation{Department of Physics, University of Basel, Klingelbergstrasse 82, CH-4056 Basel, Switzerland}
\author{Rapha\"elle Delagrange}
\affiliation{Department of Physics, University of Basel, Klingelbergstrasse 82, CH-4056 Basel, Switzerland}
\author{Roy Haller}
\affiliation{Department of Physics, University of Basel, Klingelbergstrasse 82, CH-4056 Basel, Switzerland}
\author{Lujun Wang}
\affiliation{Department of Physics, University of Basel, Klingelbergstrasse 82, CH-4056 Basel, Switzerland}
\alsoaffiliation{Swiss Nanoscience institute, University of Basel, Klingelbergstrasse 82, CH-4056 Basel, Switzerland}
\author{P\'eter Makk}
\affiliation{Department of Physics, Budapest University of Technology and Economics and Nanoelectronics Momentum Research Group of the Hungarian Academy of Sciences, Budafoki ut 8, 1111 Budapest, Hungary}
\author{Kenji Watanabe}
\affiliation{Research Center for Functional Materials, National Institute for Material Science, 1-1 Namiki, Tsukuba 305-0044, Japan}
\author{Takashi Taniguchi}
\affiliation{International Center for Materials Nanoarchitectonics, National Institute for Material Science, 1-1 Namiki, Tsukuba 305-0044, Japan}
\author{Christian Sch\"onenberger}
\affiliation{Department of Physics, University of Basel, Klingelbergstrasse 82, CH-4056 Basel, Switzerland}
\alsoaffiliation{Swiss Nanoscience institute, University of Basel, Klingelbergstrasse 82, CH-4056 Basel, Switzerland}

\title[\textsf{achemso}]
  {Compact SQUID realized in a double layer graphene heterostructure}

\keywords{Double layer graphene, SQUID, van der Waals heterostructure, helical states, supercurrent \LaTeX}

\begin{document}

\begin{abstract}
Two-dimensional systems that host one-dimensional helical states are exciting from the perspective of scalable topological quantum computation when coupled with a superconductor. Graphene is particularly promising for its high electronic quality, versatility in van der Waals heterostructures and its electron and hole-like degenerate 0$th$ Landau level. Here, we study a compact double layer graphene SQUID (superconducting quantum interference device), where the superconducting loop is reduced to the superconducting contacts, connecting two parallel graphene Josephson junctions. Despite the small size of the SQUID, it is fully tunable by independent gate control of the Fermi energies in both layers.  Furthermore, both Josephson junctions show a skewed current phase relationship, indicating the presence of superconducting modes with high transparency. In the quantum Hall regime we measure a well defined conductance plateau of 2$e^2/h$ an indicative of counter propagating edge channels in the two layers. Our work opens a way for engineering topological superconductivity by coupling helical edge states, from graphene's electron-hole degenerate 0$th$ Landau level via superconducting contacts.
\end{abstract}

The coupling of one-dimensional (1D) helical states to a superconductor is expected to generate  zero energy Majorana bound states \cite{San-Jose2015, Fu2008, Alicea2010}. Currently, the most popular approach to engineer Majorana zero modes is to use 1D semiconducting nanowires with large spin orbit coupling in contact with a superconductor exposed to large magnetic fields ($B$) \cite{Mourik2012, Das2012, Rokhinson2012, Suominen2017}. However, two-dimensional (2D) systems are required to realise scalable architectures for topological quantum computation, and some efforts have been made in this direction using semiconducting quantum well structures \cite{Hart2014,Pribiag2015,Fornieri2019, Mayer2019}. Nevertheless, these materials are difficult to fabricate and often suffer from a bulk contribution to the current transport \cite{Nowack2013, Charpentier2013}. Graphene, on the other hand, exhibits extremely high electronic quality and can be assembled into versatile heterostructures along with other van der Waals (vdW) materials to create novel device characteristics\cite{Yankowitz2019}.

Graphene, by itself, is neither a topological insulator nor possesses helical states at $B=0$. However, it can be used in vdW heterostructures to engineer helical states. Previous experiments have shown the manifestation of this state by applying a large tilted magnetic field up to 30\,T to increase the Zeeman energy\cite{Young2013}, by gating large angle twisted bilayer graphene to opposite filling factors of $\pm$1 \cite{Sanchez-Yamagishi2016}, or by using SrTiO$_3$ with its large dielectric constant as a substrate to screen the long range Coulomb interactions \cite{Veyrat2020}. All these approaches rely on the electron and hole-like, four fold (spin and valley) degenerate 0$th$ Landau level (LL) in graphene. In the quantum Hall (QH) regime, the ground state at charge neutrality is determined by the splitting of the four-fold degenerate LLs into spin or valley polarised states, which depend on the details of the interaction terms \cite{Abanin2006,Young2012,Kharitonov2012,Herbut2007, Amet2014}. The other necessary component for topological superconductivity is the coupling of the helical states to a superconductor in such a way, that an electron (hole) propagating forward in one channel is reflected as a hole (electron) propagating backwards in the other. In a system where two layers of graphene are placed one on top of the other, this can be achieved with crossed Andreev reflections (CAR) \citep{Park2019}. This has been shown to be possible at B=0 if the two layers are closer than the superconducting coherence length of the electrodes \cite{Park2019}. Combined with the possibility of inducing superconducting correlations in graphene QH edge states \cite{Amet2015, Lee2017}, an engineered topological superconductivity should be experimentally accessible.

\begin{figure}[htbp]
	\centering
	\hspace*{0cm}
	\includegraphics[width=8.9cm]{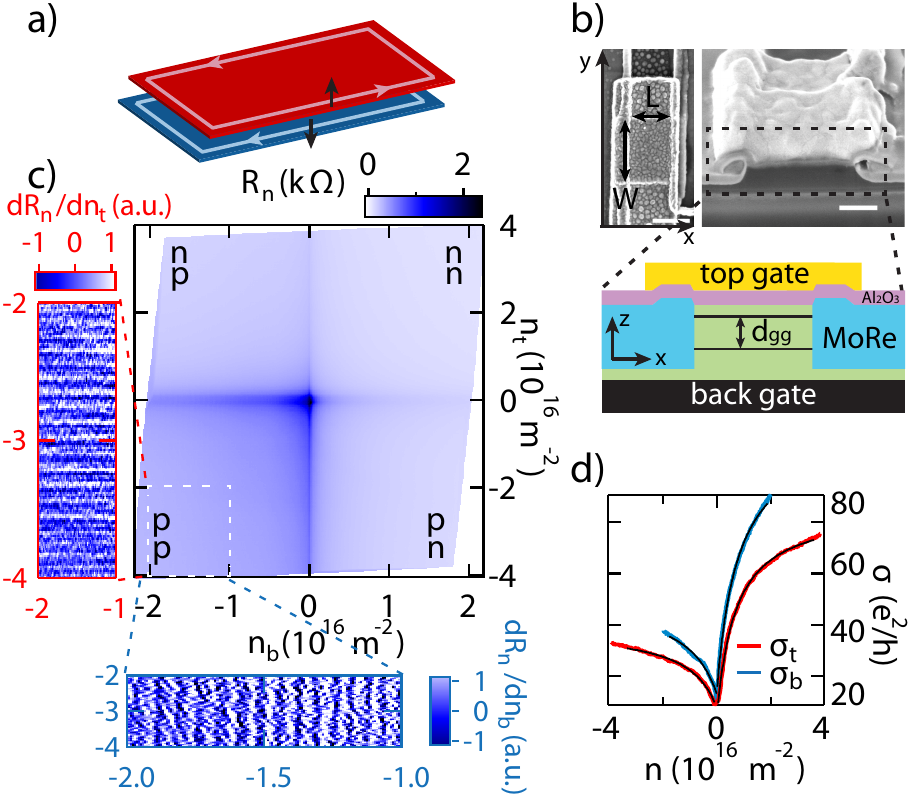}	
	\caption{a) Schematic of a DLG device in the QH regime, where the top (bottom) graphene layer is indicated in red (blue) and at a filling factor of +1 (-1). The edges states are spin polarized and together form a helical QH state. b) SEM images of a DLG JJ with a schematic of the cross-section. The image on the top left shows the top view of J$_1$ lying in the xy-plane with a length of 580\,nm and a width 1\,$\mu$m.  Scale bar is 500\,nm. On the right  a cross-section of the J$_1$ with a $d_{gg}$ of 12\,nm. The sample is cut parallel to the x axis. Scale bar represents 200\,nm. The schematics on the bottom shows a sketch of the cross-section. The thin (thick) black lines corresponds to the graphene (graphite gate), the green areas to the hBN, blue to the MoRe contacts, pink to the Al$_2$O$_3$ and yellow to the gold top gate. c) Normal state resistance ($R_n$) of J$_2$ as a function of top and bottom carrier density. The red (blue) framed graph shows the derivative of $R_n$ in the white dashed enclosed region with respect to $n_t$ ($n_b$).  d) Cuts of the $R_n$ measurements at $n_t=0$ (blue) and $n_b=0$ (red) are shown. The black lines are fits of the conductivity as a function of $n$ given by the mobility and contact resistance.}
	\label{fig:Rn}
\end{figure}

In this work, we investigate Josephson junction made from a vdW heterostructure with two graphene layers separated by a thin hexagonal boron nitride (hBN) crystal. In the QH regime, we gate the two layers to the filling factors of $\pm$1, which are states  with opposite spin and propagation direction\cite{Sanchez-Yamagishi2016} (see Fig.\ref{fig:Rn}\,a). This mimics an artificial 1D helical conductor and manifests as a well defined conductance plateau of 2$e^2/h$, where $h$ is the Planck constant and $e$ the elementary charge. Superconducting molybdenum-rhenium (MoRe) side contacts, with a critical magnetic field $B_c$ of $\sim$8\,T, connect the double layer graphene (DLG) forming two parallel MoRe-graphene-MoRe Josephson junctions (JJ) as shown in Fig.\ref{fig:Rn}\,b. As a whole, it constitutes a compact SQUID whose superconducting loop is reduced to the electrodes. The critical current ($I_c$) in both JJs can be independently tuned to define a symmetric or an asymmetric SQUID. In the later case, we measure the current phase relation (CPR) of both graphene JJs and observe superconducting modes with a high transparency. Demonstrating helical edge modes in two graphene layers placed within a superconducting coherence length of each other represents our first step towards engineering topological superconductivity in graphene based heterostructures.


We fabricated six-layered vdW heterostructures shown in Fig.\ref{fig:Rn}\,b by a standard polycarbonate assisted pick-up technique \cite{Zomer2014}. The hBN between the graphene sheets, has three atomically defined thicknesses - 12\,nm, 25\,nm and 50\,nm for junctions J$_1$, J$_2$ and J$_3$ respectively. The self-aligned MoRe side contacts were sputtered after etching the contact region with CHF$_3$/O$_2$ \cite{Wang2013}. The hBN on the bottom and top were used as gate dielectrics. Aluminium oxide (Al$_2$O$_3$) with a thickness of 30\,nm was grown by atomic layer deposition to electrically insulate the top gate from the contacts as well as from the etched edges of the stack, which define the mesa. A scanning electron microscope image of the top view and a cut made by a focused ion beam are shown in Fig.\ref{fig:Rn}\,b. The fabrication details can be found in the SI.  

First we characterise the junctions by measuring the normal state resistance ($R_n$), which is shown for J$_2$, in Fig.\ref{fig:Rn}\,c and for all three junctions in the SI. It was measured at a constant voltage bias of 4\,mV, which is well above 2$\Delta$, where $\Delta$=1.3\,meV is the superconducting gap of MoRe \cite{Indolese2018}.  At this voltage bias, the current is carried by quasi particles and any influence of the superconducting contacts can be neglected. The electron densities of the top ($n_t$) and the bottom ($n_b$) layer are calculated by an electrostatic model for each pair of top ($V_{tg}$) and back gate ($V_{bg}$) voltage (see SI). Two clear lines of enhanced resistance at $n_t$ and $n_b$ equal to zero, correspond to the Dirac points (DP) of the individual layers, and split the $R_n(n_b,n_t)$ into four quadrants. We observe the highest mean resistance in the quadrant, where both layers are hole (p) doped, resulting from the workfunction mismatch between graphene and MoRe, which electron (n) dopes the graphene near the contacts\cite{Calado2015}. This pn junction at both electrodes leads to a charge carrier dependent reflection probability, i.e. Fabry-P\'erot oscillations, if the phase coherence length is larger than the junction length ($L$)\cite{Young2009, Calado2015}. For a higher visibility of the oscillations, the derivatives of $R_n$ with respect to $n_b$ and to $n_t$ are shown in Fig.\ref{fig:Rn}\,c. The mobility ($\mu$) and the contact resistance ($R_c$) are determined from the charge carrier dependent conductivity $\sigma$ for $n_t=0$ and $n_b=0$ (see SI). The relatively high $\mu$ values up to 53'000\,cm$^2$/Vs and the observation of Fabry-P\'erot oscillations  indicate ballistic transport in both junctions. 

\begin{figure}[htbp]
	\centering
	\includegraphics[width=8.9cm]{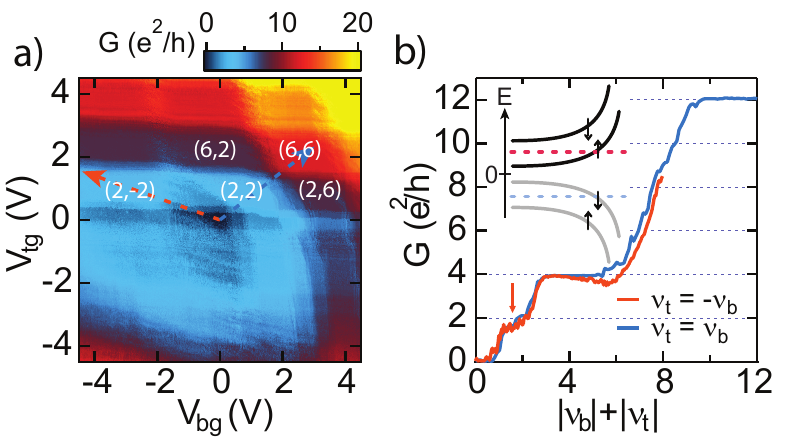}	
	\caption{a) Conductance as a function of top and back gate. Clear plateaus of constant conductance and a degeneracy lifting of the 0$th$ Landau level are observed. The white numbers indicate some filling factors in the top (left number) and bottom (right number) graphene layer for the conductance plateau. b) Cuts of the conductance map shown in a) along equal $n_b$=$n_t$ (blue) and opposite $n_b$=$-n_t$ with $n_t>0$ (orange) as a function of the sum of the absolute filling factors. Inset: Energy spectrum of the 0$th$ Landau level for a single layer graphene with spin and valley degeneracy lifted. The black (gray) lines correspond to electron (hole)  like states. The chemical potential of the top graphene (red dashed line) is set between the filling factors of $\nu=1$ and 2. The blue dashed line indicates the position of the chemical potential of the bottom graphene set between $\nu=-1$ and -2. This situation corresponds to the conductance plateau pointed at by the orange arrow.}
	\label{fig:hQH}
\end{figure}

To explore the presence of the engineered helical state in our junctions, we measure the device conductance ($G$) in an out-of-plane $B$ of 5\,T (see Fig.\ref{fig:hQH}\,a). We observe well defined conductance plateaus as a function of $V_{tg}$ and $V_{bg}$. In the QH regime the current is carried by a discrete number of edge channels, which is given by the filling factor $\nu$ \cite{Zhang2005}. For electrically separated DLG, there is no coupling of edge channels, and $G$ of each plateau is given by the sum of the absolute filling factors in both layers, $G=\frac{e^2}{h}(|\nu_b|+|\nu_t|)$ with $\nu_i=\pm 2,\pm6,\pm10,...$, where $i$ is the index for the bottom or the top layer, assuming that the four fold degeneracy of each LL is not lifted. 

For equal charge carrier density in both layers, one expects a sequence of $\nu_{tot}=|\nu_b|+|\nu_t|=4,12,20,...$, when neither of the degeneracy, i.e. spin or valley, is lifted. In Fig.\ref{fig:hQH}\,b the blue curve corresponds to the cut along $\nu_t=\nu_b$, where the edge channels in the two graphene layers are electron like and propagate in the same direction. We observe plateaus with  $4\frac{e^2}{h}$, $12\frac{e^2}{h}$ and $20\frac{e^2}{h}$ as expected. Exceptionally, an additional plateau with $G$=$2\frac{e^2}{h}$ exists, which appears due to the lifting of the degeneracies of the 0$th$ LL by many-body correlation \cite{Zhang2006}.  The same value in conductance is also observed for $\nu_t=-\nu_b$ with $\nu_t>0$. In this case the n-doped top layer and the p-doped bottom layer host counter propagating edge channels. The presence of an insulating ground state at charge neutrality indicates the lifting of the valley and spin degeneracy such that edge channels for $\nu=1$ and $\nu=-1$ must be spin polarized with opposite spins, and form a 1D helical state \citep{Sanchez-Yamagishi2016}. While the MoRe is still superconducting at 5\,T, we do not observe superconducting pockets\cite{Amet2015} for $\nu_{t/b}=\pm1$, even when the ac-current modulation was reduced to 50\,pA.

\begin{figure}[htbp]
	\centering
	\hspace*{0cm}
	\includegraphics[width=8.9cm]{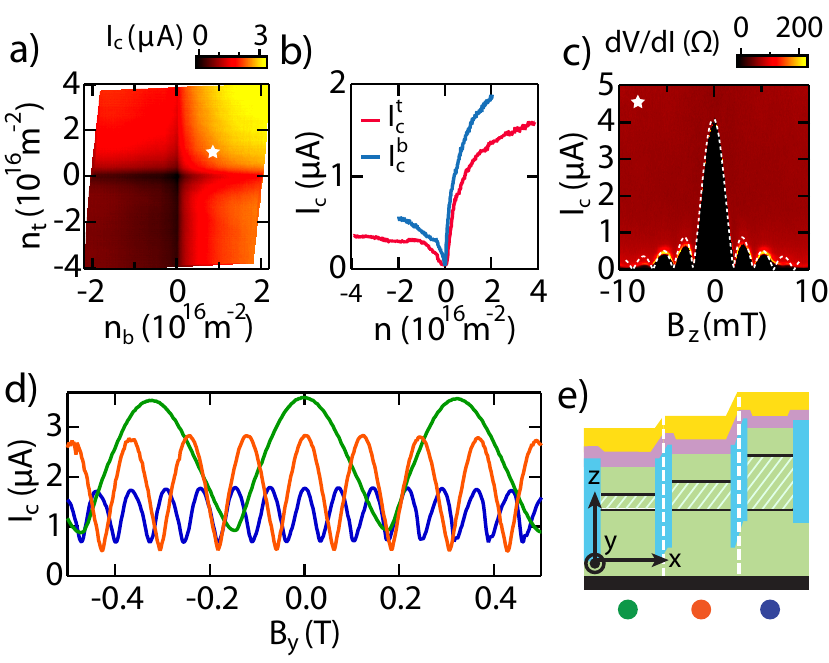}	
	\caption{a) $I_c$ as a function of $n_t$ and $n_b$. b) Cuts along $n_t=0$ (blue) and $n_b=0$ (red). c)Interference pattern of the DLG JJ as a function of out-of-plane magnetic field $B_z$. The white star in Fig.\ref{fig:Ic}\,a indicates the position in $n_b$ and $n_t$, where the interference pattern is measured. The interference pattern is overlaid with a Fraunhofer interference pattern (white dashed line) given by Eq.\ref{Eq:FH}. d) Interference pattern of DLG JJ as a function of in-plane magnetic field $B_y$. The different colors correspond to different $d_{gg}$. e) Schematic cross-section of the different JJ. The coloured dots are related to the interference pattern in d). The flux through the striped areas has to be considered to calculate the period in magnetic field.}
	\label{fig:Ic}
\end{figure}

Next, we characterise the superconducting properties of our junctions at $B=0$, which is shown in Fig.\ref{fig:Ic} for J$_2$. Here, $I_c$ was measured as a function of $n_t$ and $n_b$ (see Fig.\ref{fig:Ic}\,a) by the appearance of a non-zero trigger voltage ($V_{trig}$) over the junction, while sweeping the bias current (as detailed in SI). The observed smallest $I_c$ is of around 40\,nA at $n_t=n_b=0$.  The critical current of the bottom (top) layer $I_c^{b}$ ($I_c^{t}$) extracted at $n_t=0$ ($n_b=0$) shows a similar dependence of $n_b$ ($n_t$) as the conductivity (see Fig.\ref{fig:Ic}\,b), as expected from the Ambegaokar-Baratoff relation ($R_nI_c\propto \Delta$)\cite{Ambegaokar1963}.  Here, the superconducting coherence length in graphene $\xi_g=\frac{\hbar v_F}{\pi \Delta}=160$\,nm, where $\hbar$ is the reduced Plank constant and $v_F$ is the Fermi velocity of graphene, is smaller then $L$. For such a long junction ($\xi_g<L$) the product of $R_n$ and $I_c$ is proportional to the Thouless energy\cite{Borzenets2016} instead of $\Delta$ (see SI). 

A magnetic flux $\Phi_z=\Phi(B_z)$, induced by an out-of plane magnetic field $B_z$ threading through a planar JJ creates a phase difference between the superconducting trajectories. That results in an interference pattern, which is measured as the dependence of $I_c$ with respect to  $B_z$, obtained by sweeping a dc-current with an added ac-modulation of 10\,nA and detecting the differential resistance (see Fig.\ref{fig:Ic}\,c). If the current distribution is homogeneous along the y-axis, $I_c$ can be expressed as a function of $B_z$ as \cite{Tinkham1996},
 
 \begin{equation}
 	\label{Eq:FH}
	I_{c}(B_{z})=I_{c}(0)\left|\frac{\sin(\pi\Phi(B_z)/\Phi_{0})}{\pi\Phi(B_z)/\Phi_{0}}\right|,
 \end{equation}
where $\Phi_0$ is the magnetic flux quantum given by $h/2e$ and $\Phi_z=W(L+2\lambda)B_z$ \cite{Tinkham1996} with $\lambda \approx 150$\,nm equal to the London penetration depth of MoRe. Since the measured $I_c$ describes a Fraunhofer like interference pattern, we conclude that the supercurrent density is indeed homogeneous in both layers. The period of the oscillation in $B_z$  matches the sample dimensions. Note, that $I_c(B_z)$ drops to zero within a small field of 10\,mT.

The measurements shown in Fig.\ref{fig:Ic}\,a $\&$ c can not distinguish between the current that flows  in the top or the bottom layer of the DLG JJs. In contrast, the in-plane magnetic field $B_y$ (y-axis) dependence of $I_c$ is sensitive to the relative magnitude of $I_c$ in both junctions. For this measurement, the alignment of $B$ is crucial (see SI), since a small out-of-plane component leads to a fast decay of $I_c$ as we have seen before (see Fig.\ref{fig:Ic}\,c).  The magnetic flux $\Phi_y$, given by $B_y-$ threading through the loop formed by the two vertically stacked graphene layers and the superconducting electrodes, induces a phase difference between the two JJs, given by $\varphi_t=\varphi_b+\frac{2 \pi \Phi_y}{\Phi_0}$, where $\varphi_i$ is the phase difference over the i-th JJ. In general, the total supercurrent ($I_s$) of the SQUID is given by 
\begin{equation}
	\label{Eq:Ictot}
	I_s(\varphi_t,\varphi_b)=I_c^t g^t(\varphi_t)+ I_c^b g^b(\varphi_b),
\end{equation}
where $g^i$ is the CPR of the i-th JJ and thereby an odd 2$\pi$ periodic function. In the case of  JJ with a tunnel barrier as a weak link, $g(\varphi)=\sin(\varphi)$. As mentioned, $\varphi_t$ can be replaced by $\varphi_b+\frac{2\pi \Phi_y}{\Phi_0}$ in Eq.\ref{Eq:Ictot}. The critical current as a function of $\Phi_y$ is then obtained by maximizing $I_c(\varphi_b,\Phi_y)$ over $\varphi_b$ for a given $B_y$.
\begin{equation}
	\label{Eq:Ic}
	I_c(\Phi_y)=\max_{\varphi_b} \left\lbrace I_c^t g^t\left(\varphi_b+\frac{2\pi\Phi_y}{\Phi_0}\right)+ I_c^b g^b(\varphi_b)\right\rbrace,
\end{equation}


The measurements for all three SQUIDs show an expected $\lvert\cos\left(\pi \Phi_y/\Phi_0\right)\rvert$ like behavior (see Fig.\ref{fig:Ic}\,d), for a symmetric SQUID configuration, i.e. when $I_c^t\approx I_c^b$\cite{Angers2008} and an assumed sinusoidal CPR. From the periodicity of the interference pattern in $B_y$ we extract the cross sectional area $L \times d_{gg}$, for which we obtain a good agreement with their physical dimensions [J$_1$: (580\,nm$\times$12\,nm, J$_2$: (650\,nm$\times$25\,nm, J$_1$: (530\,nm$\times$50\,nm)]. Note, that the minimal value of $I_c$ never reaches zero, revealing either a small difference between the critical currents of both layers or a non-sinusoidal CPR of the junctions. In the SI we show that even for $I_c^t=I_c^b$ the total $I_c$ does not go to zero.  We observe no signs of a CAR contribution with a period of $h/e$, which would alter the value of every second maxima, although the $d_{gg}$ is on the order of $\xi_{MoRe}$ \cite{Talvacchio1986}. Another possibility is the momentum mismatch between the two graphene layers by a twist angle \cite{Sanchez-Yamagishi2016}. 

Finally we characterise each JJ by measuring the CPR employing the high individual tunability of $I_c^t\,\&\, I_c^b$.  The CPR gives an insight into the Cooper pair transport across the junction, e.g. transparency of Andreev bound states. For $I_c^b$ much larger than $I_c^t$ ($\frac{I_c^b}{I_c^t} \geq 10$), $\varphi_b$ can be assumed constant in Eq.\ref{Eq:Ic} such that $f^b$ stays at its maximum value. This results in $I_c=I_c^b+I_c^t g^t(\varphi_b^{max}-\frac{2\pi \Phi}{\Phi_0})$ \cite{DellaRocca2007}. In this case a change in flux leads to an oscillation around $I_c^b$ with an amplitude of $\pm I_c^t$, while the shape of the oscillation is given by $g^t$. Hence, by using an asymmetric configuration the CPR can be obtained.

For the sake of clarity, we discuss the CPR for a JJ in the short junction limit. Here, the CPR is determined by the number of transport channels and their transparencies. The critical current of a short JJ is then given by
\begin{equation}
	\label{Eq:t}
	I_c=\frac{e\Delta}{2\hbar}\sum_n \frac{t_n \sin(\varphi)}{\sqrt{1-t_n \sin^2(\varphi/2)}},
\end{equation}
where $t_n$ is the transparency of the n-th channel\cite{Kulik1975, Golubov2004, Spanton2017}. For a superconducting tunnel junction, one expects a sinusoidal CPR since only channels with low transparency ($t_n \ll 1$) contribute to  transport. If the channel transparencies increase the maximum of the CPR starts to deviate from $\varphi_{max}=\pi/2$ towards $\pi$\cite{Thompson2017,Bretheau2017a, Nanda2017, English2016}. This deviation is quantified by the skewness of the CPR, defined as $S=\frac{\varphi^{max}-\pi/2}{\pi/2}$ \cite{Nanda2017}. The skewness is 0, when the CPR is sinusoidal ($t_n \ll 1$), and is 1 when the CPR is sawtooth like ($t_n=1$). Note, that $S$ and $t_n$ have a non-linear relation, such that a single channel with $t$=0.9 causes only a skewness of 0.35.

\begin{figure}[htbp]
	\centering
	\hspace*{-0.3cm}
	\includegraphics[width=8.4cm]{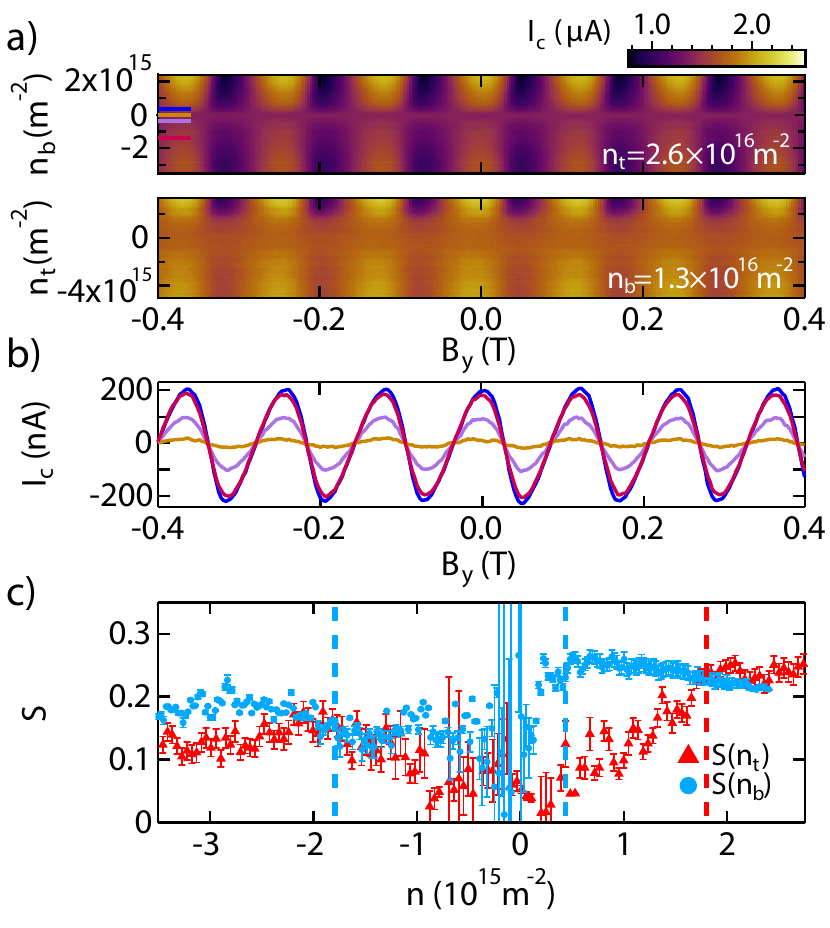}	
	\caption{a)Interference pattern as a function of $n_b$ (top) respectively $n_t$ (bottom) and in-plane magnetic field for fixed top respectively bottom charge carrier density. b) Cuts of the interference pattern as a function of $B_y$ for different $n_b$ after subtracting the critical current of the reference junction. The color of the curves corresponds to the color of the bars at the left axis of the top graph in a), which indicates the positions in $n_b$ where the line cuts are taken. c) The skewness of the CPR as a function of $n_b$ (blue) and $n_t$ (red) are shown. The blue (red) dashed line corresponds to a ratio of $I_c^b/I_c^t$ ($I_c^t/I_c^b$) equal to 0.15.}
	\label{fig:Bin}
\end{figure}

In Fig.\ref{fig:Bin}\,a we show the results of the CPR measurements around the DPs. The reference junction was highly n doped and has a large $I_c$ ($I_c^t \approx 1.5\,\mu A\,\&\,I_c^b\approx 2\,\mu A$). Four examples of the CPR at different densities are shown in Fig.\ref{fig:Bin}\,b. The positions in $n_b$ are indicated in Fig.\ref{fig:Bin}\,a. A strong decrease of the amplitude at the DP is observed (orange). For values of $n_b \neq 0$ we observe a skewed sinusoidal like oscillation. To extract the skewness, the measurement data were fitted over six periods by using,

\begin{equation}
	 \label{Eq:Skew}
	 I_c^i= \sum_{n=1}^5 a_n\sin(nf\times(B_y+B_0)),
\end{equation}
where $a_n$ is the pre-factor of the n-th harmonic, $f$ the frequency of the oscillations, and $B_0$ the shift in $B_y$ with respect to the first zero crossing of the oscillation. The pre-factors and their ratio is shown in the SI. The skewness, extracted from the maximum of the fit at each $n_{t/b}$, is plotted in Fig.\ref{fig:Bin}\,c. The largest $S\approx0.25$ was observed when the layers are n doped. For the p doped graphene a $S$ of around 0.15 was deduced from the data. By assuming a single channel in the short junction limit, these values would correspond to a transparency of $t=0.7$ for $S=0.25$ and $t=0.6$ for $S=0.15$ in Eq.\ref{Eq:t}. Towards the DPs we observe a reduction of the skewness, which indicates that the transparency of the modes reduces when compared to higher doping. $S(n_b)$ also appears to decrease at large electron density, but is possibly due to the assumption $\frac{I_c^b}{I_c^t} \gg 1$ breaking down. The interference signal rapidly diminishes for densities lower than $2\times 10^{14}$\,m$^{-2}$. This corresponds to a wavelength $\lambda_F \approx 250\,$nm comparable to the length of the junction. For L<$\lambda_F$, the remaining modes have an angle and will be suppressed at the pn-junction that appear close to the contacts. This explains the rapid reduction in the $I_c$ and the sinusoidal CPR close to the DPs.
 
We note, that the self inductance $L_s$ of the SQUID can also lead to a non-linear dependence of $\varphi$ as a function of $B$ caused by screening of the external magnetic field by the current in the SQUID loop \cite{Golubov2004}. This effect is especially dominant around $\Phi=\Phi_0/2$, where the current in the loop is the largest and can mimic a non-sinusoidal CPR. However, in our compact DLG SQUID, the weak links (i.e. the graphene layers) are intersected by two short superconducting segments. This results in a negligible contribution from the kinetic inductance irrespective of the superconducting material and also a minimal geometrical inductance (see SI), in contrast to a regular SQUID geometry.

In conclusion, we engineer a helical state in a double layer graphene van der Waals heterostructure using the unique properties of the 0$th$ LL of graphene. This is reflected by the observation of a well defined conductance plateau of 2$e^2/h$, when each of the two layers hosts an opposite chirality edge channel, i.e. for $\nu_t=-\nu_b$ with $|\nu_t|=1$. We show that the supercurrent is carried by both layers and the critical currents are individually tunable and show a strong carrier density dependence. The measurement of the current phase relation reveals the high transparency in both graphene Josephson junctions. Further, the loop inductance of such a compact SQUID is negligible in interference measurements. We do not observe crossed Andreev reflection between the two graphene layers possibly due to the short coherence length of MoRe or a momentum mismatch between the two misaligned layers. Our graphene based heterostructure, however, allows a small interlayer spacing and the alignment of the crystal axes of the graphene layers by the tear and stack technique \cite{Kim2016a}, which paves the way for future experiments to couple the helical quantum Hall edge states by crossed Andreev reflection.

\begin{acknowledgement}

The authors acknowledge the support by the Swiss Nanoscience Institute (SNI), the ERC project TopSupra (787414), the European Union Horizon 2020 research and innovation program under grant agreement No. 785219 (Graphene Flagship), the Swiss National Science Foundation and the Swiss NCCR QSIT. A.K. was supported by the Georg H. Endress foundation. Topograph FlagERA network, OTKA FK-123894. This research was supported by the National Research, Development and Innovation Fund of Hungary within the Quantum Technology National Excellence Program (Project Nr. 2017-1.2.1-NKP-2017-00001). P.M. acknowledges support from the Marie Curie and Bolyai fellowships. K.W. and T.T. acknowledge support from the Elemental Strategy Initiative conducted by the MEXT, Japan ,Grant Number JPMXP0112101001,  JSPS KAKENHI Grant Numbers JP20H00354 and the CREST(JPMJCR15F3), JST. D.I. acknowledges the fruitful discussions with Simon Zihlmann.

\end{acknowledgement}

\section*{Author contributions}
D.I, P.M, R.D. and C.S. conceived the experiment. D.I. did the fabrication of the device with the input of L.W. The measurements were performed by D.I. with the input of P.K and A.K. The hBN crystals was provided K.W. and T.T. R.H. maintained the measurement setup and improved it together with D.I. Further D.I. analysed the data with inputs of P.K. and A.K. D.I. and P.K. wrote the manuscript with contribution from all authors.

\begin{suppinfo}

The following files are available free of charge.
\begin{itemize}
  \item Supporting Informations: Compact SQUID realized in a double layer graphene heterostructure: Detailed description of device fabrication, experimental methods and complementary measurements
  \item All measurement data in this publication are available in numerical form (DOI:10.5281/zenodo.3886697).
\end{itemize}

\end{suppinfo}

\bibliography{refs}

\end{document}





\subsection{Device fabrication}
The fabrication of the van der Waals heterostructure (vdWh), superconducting contacts, and gates  is described in the following section. In a first step all the 2D materials, i.e. graphite and hBN, were exfoliated on a silicon wafer with an oxide thickness of 295\,nm using the low adhesion tape ELP BT-150P-LC supplied by Nitto. We identified the graphene by the optical contrast difference of 4\,$\%$ in the green channel with respect to the substrate \cite{Ni2007}. The thickness of the middle hBN and its plateaus were measured by an atomic force microscope (AFM) using a Bruker Dimension 3100 in tapping mode at ambient conditions. 

The stacking of the different crystals was done by a well established technique, which uses a polycarbonate (PC) film on a polydimethylsiloxane (PDMS) pillow \cite{Zomer2014}. Here, the such fabricated vdWh consists from bottom to top out of thick graphite, a bottom hBN, a bottom graphene, a middle hBN, a top graphene, and a top hBN flake. Important to mention is that only the top surface of the top hBN flake is in contact with the PC, such that the entire stacking process is fully assisted by the van der Waals forces and it can be considered dry and polymer free. Further the encapsulation protects the graphene layers from contaminations during the device fabrication. In the end the stack was placed at 170\,$^\circ$C on a intrinsic silicon wafer with an oxide thickness of 285\,nm. At these temperature the PC detaches fully from the PDMS pillow. The PC residues on top of the stack can be dissolved in dichlormethane for 1\,h at room temperature. To clean the top hBN's surface, the stack was annealed at a temperature of 300\,$^\circ$C for 3\,h in forming gas (H$_2$ 8$\%$/N$_2$ 92$\%$) at a background pressure of 20\,mbar. This removes remaining PC residues. 

During the stacking process it is possible that impurities are trapped between two layers, which manifest themselves in the appearance of bubbles. Using an AFM we can map the clean areas, i.e. the areas without bubbles, which are used for the device fabrication. Further we can determine the thickness of the different hBN, which is important to determine the exact etching times and for the calculation of the electrostatics.

The two layers of graphene have been connected by several common superconducting edge contacts forming several Josephson junctions (JJs) in series. To fabricate the contacts we used standard electron beam lithography (EBL). A e-beam resist of PMMA 950k, which is dissolved in anisole (concentration of 5.5$\%$), was spin coated with 4000\,rpm for 40\,s on the sample, resulting in a film thickness of 380\,nm. The EBL was performed with an acceleration voltage of 20\,kV and a dose of 400\,$\mu$C/cm$^2$. The resist was developed for 1\,min in a IPA/deionized water mixture (7/3) cooled to $\sim{}$5$^\circ$C and was then blow dried with nitrogen. To fabricate the superconducting edge contacts the stack was etched by a reactive ion etching using a CHF$_3$/O$_2$ plasma with 40\,sccm/4\,sccm at a background pressure of 60\,mTorr and a power of 60\,W. The rate of the etching recipe was calibrated in advance to have a very precise control of the amount of etched hBN. This is crucial, since one has to stop the etching process in the bottom hBN layer, such that the bottom gate is electrically insulated from the MoRe electrodes, but both graphene layers can be contacted simultaneously. After the etching, the MoRe was sputtered in a AJA ATC Orion using still the same PMMA mask. For the sputtering we used a target of Mo/Re 1:1, a power of 100\,W, a background pressure of 2\,mTorr, and a constant Argon flow of 30\,sccm. The contacts have a thickness of 80\,nm. The lift-off was done in acetone at 50\,$^\circ$C. In a next step the MoRe was contacted by Cr/Au (5\,nm/125\,nm) using another EBL defined mask and electron beam evaporation at a pressure of 5e$^{-7}$mbar. After the lift-off, an etching mask was prepared by EBL to shape the mesa. To insulate the structure from the topgate it was overgrown by an uniform Al$_2$O$_3$ layer of 30\,nm using atomic layer deposition (ALD), which  involved trimethylaluminium (TMA) and water. We observed that for a homogeneous growth of the Al$_2$O$_3$ on the vdWh a short O$_2$ plasma (flow 16\,sccm, pressure 250\,mTorr, power 30\,W, time 20\,s) is needed. This removes remaining polymer residues from the fabrication and leads to a homogeneous wetting of the surfaces. In last step we deposited the metallic topgate.

\subsection{Normal state resitance}
\begin{figure}
	\centering	
	\includegraphics[scale=0.9]{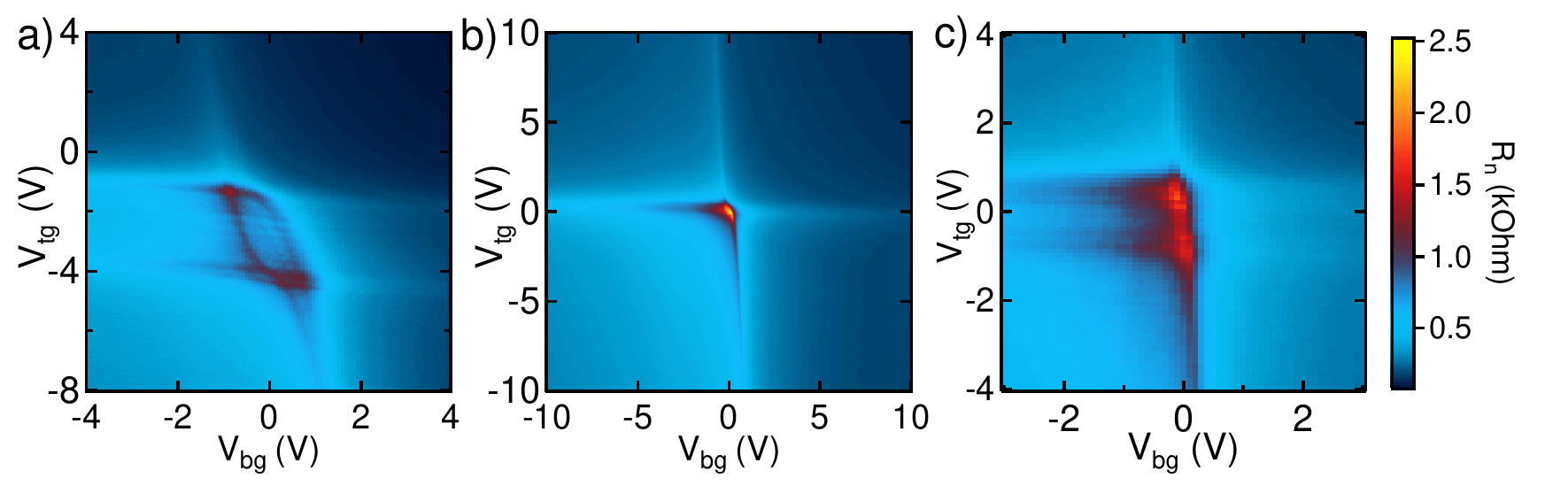}
	\caption{a) Normal state resistance as a function of top and bottom gate for $d_{gg}$=12\,nm. b) Normal state resistance as a function of top and bottom gate for $d_{gg}$=25\,nm. c) Normal state resistance as a function of top and bottom gate for $d_{gg}$=50\,nm.}
	\label{fig:Rn_all}
\end{figure}

The normal state resistance ($R_n$) was measured for three different junctions with different inter layer distances ($d_{gg}$) by applying a bias voltage of 4\,mV, which is larger then twice $\Delta_{MoRe}=1.3$\,meV \cite{Indolese2018}. For the junctions with $d_{gg}$=12\,nm and 50\,nm we observed two Dirac points (DP) as a function of the top gate voltage ($V_{tg}$) (see Fig.\ref{fig:Rn_all}\,a and c). We attribute this behavior to a inhomogeneous lateral residual doping in the top graphene layer. The splitting of the DP as a function of the back gate voltage ($V_{bg}$) around charge neutrality (see Fig.\ref{fig:Rn_all}\,a) can then be explained by the density of states (DOS) dependent screening of the top gate by the different top graphene regions. For the junction with a thickness of 25\,nm, this behavior is less pronounced, but the DP of the top graphene is broadened in charge carrier density compared to the bottom one, which can may be attributed to the same effect.

The Fabry-P\'erot cavity length ($L_c$), i.e. the size of the p doped region, at large $n_t$ and $n_b$ is extracted from the location of neighbouring resistance maxima in charge carrier density using Eq.\ref{Eq:FPosc} \cite{Handschin2017}. 

\begin{equation}
	\label{Eq:FPosc}
	L_c=\frac{\sqrt{\pi}}{\sqrt{n_{i+1}}-\sqrt{n_i}},
\end{equation}
where $n_i$ is the position in carrier density of the i-th peak in resistance. We obtain a length of around 550\,nm. The comparison to $L$=650\,nm of J$_2$ indicates that the n doped region at each contact is of the order of 50\,nm for large densities.

To extract the mobility $\mu$ and contact resistance $R_c$ of J$_2$ we fit the conductivities, which are plotted in the article in Fig.1\,d, by:

\begin{equation}
\label{Eq:Mobility}
\sigma^{-1}=\frac{1}{e\mu n+\sigma_{0}}+\rho_{c},
\end{equation}
where $\sigma_0$ is the residual conductivity at the DP, and $\rho_c$ is the contact resistivity. From the fit we obtain an electron mobility of around 53'000\,cm$^2$/Vs (33'000\,cm$^2$/Vs) and a hole mobility of around 27'000\,cm$^2$/Vs (14'000\,cm$^2$/Vs) for the bottom (top) graphene. An $R_c$ of 170\,$\Omega$ (190\,$\Omega$) and 440\,$\Omega$ (490\,$\Omega$) is extracted for the bottom (top) graphene for the n-doping and the p-doping, respectively.

\subsection{Electrostatic model}
\label{sec:EM}
\begin{figure}[htb]
	\centering
	\includegraphics[width=10cm]{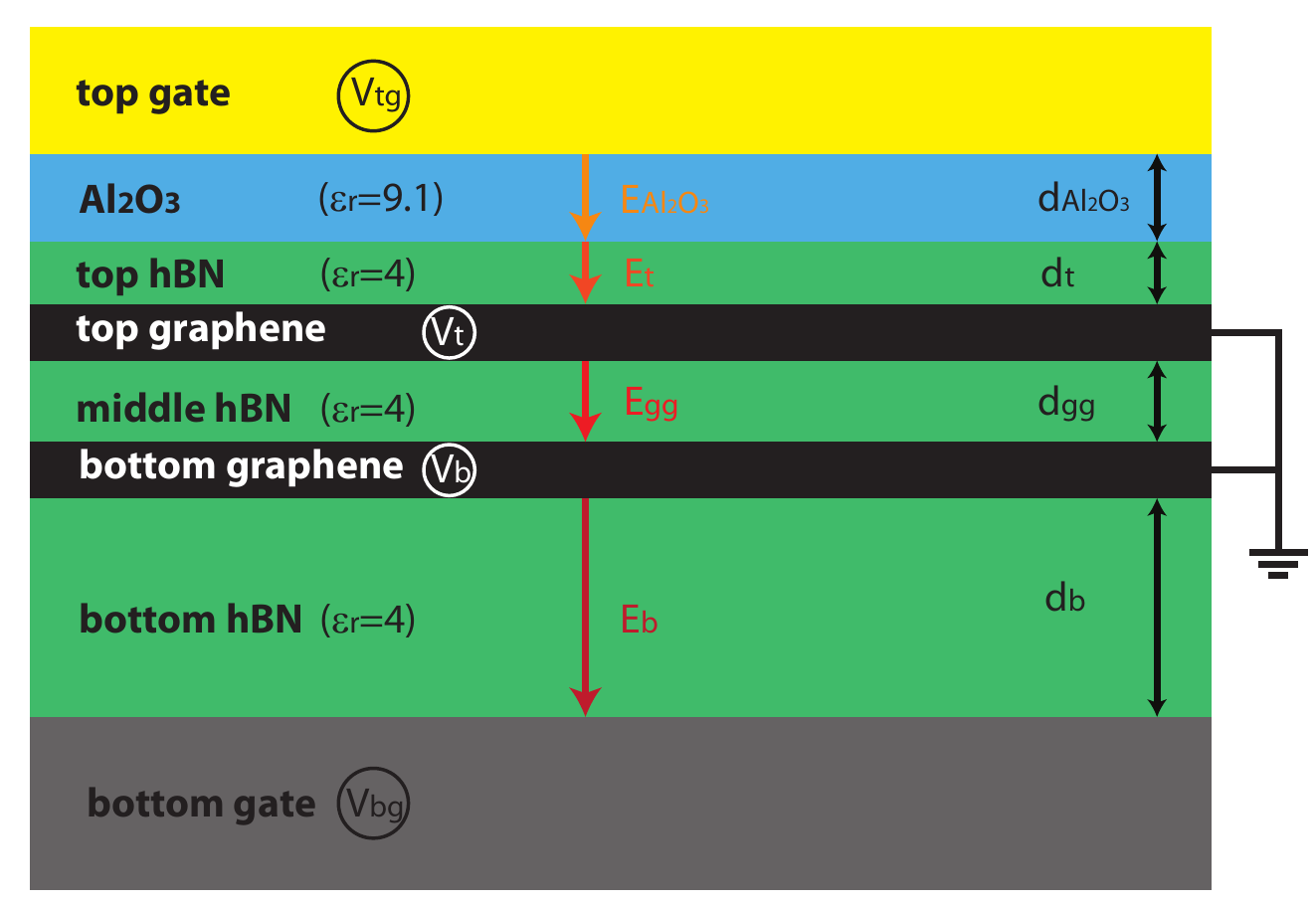}
	\caption{Schematic side view of the DLG stack. The different layers are labelled with its dielectric constant, electric potential and thickness. The arrows define the direction of the electric fields, which was taken for the calculation.}
	\label{fig:Scheme}
\end{figure}

The charge carrier density in the top ($n_t$) and the bottom ($n_b$) graphene were calculated from $V_{tg}$ and $V_{bg}$ using the electrostatic model described in the following section. The structure, which we consider is a DLG (see Fig.\ref{fig:Scheme}), consisting of the following layers listed from bottom to top: 1) graphite bottom gate 2) bottom hBN 3) bottom graphene layer 4) middle hBN 5) top graphene layer 6) top hBN 7) aluminium oxide 8) metal top gate.

The top gate is electrically separated from the top graphene by an aluminium oxide layer with a thickness of $d_{Al_2O_3}$ and a dielectric constant $\epsilon_r^{Al_2O_3}=9.1$ and the top hBN with a thickness of $d_t$ and $\epsilon_r^{hBN}=4$. A hBN with a thickness of $d_{gg}$ between the two graphene sheets electrically disconnects the two layers, which are shorted at two common 1D edge contacts. hBN was also used as a dielectric material between the bottom graphene plus electrodes and the bottom gate. The thickness of the bottom hBN layer is given by $d_b$. Since the two graphene layers are electrically shorted, they are at the same electro-chemical potential, which is chosen to be equal to zero, i.e. ground, for the following calculation. From this follows that,

\begin{equation}
\label{Eq:el-chem}
\mu_{c}^{t}-eV_t=\mu_{c}^{b}-eV_b=0,
\end{equation}

where $\mu_c^t$, $\mu_c^b$ are the chemical potential and $V_t$, $V_b$ are the electrostatic potential of the top, respectively the bottom graphene and $e$ is the elementary charge. For graphene the chemical potential is given as,

\begin{equation}
\label{Eq:chem}
\mu_{c}^i= \sgn(n_i) \hbar v_F \sqrt{\pi |n_i|},
\end{equation}
where $n_i$ the charge carrier density in the $i$-th graphene layer. The $\sgn(n_i)$ function is such that it is positive for electron doped and negative for hole doped graphene.

To describe the electrostatic situation we look carefully at the electric fields $E_i$, where the index $i$ denotes the different dielectrics, which are a consequence of applied gate voltages, quantum capacitance and charge carrier density on either graphene. The electric fields are defined as shown in Fig.\ref{fig:Scheme}.
In a first step we express $E_{Al_2O_3}$ in terms of $E_t$. If we consider the interface between the two dielectric materials to be charge free, it follows directly from the Maxwell equations that normal components of the two electric fields times their dielectric constant have to be the same at the interface. In this case $E_{Al_2O_3}$ is given by,

\begin{equation}
\label{Eq:Ealu}
E_{Al_2O_3}= \frac{\epsilon_{r}^{hBN}}{\epsilon_{r}^{Al_2O_3}}E_t.
\end{equation}

Using Gauss law we can write down $n_t$ and $n_b$ as a function of the electric fields.

\begin{equation}
\label{Eq:nt}
-e n_t=\epsilon_0\epsilon_r^{hBN}(E_{gg}-E_t)
\end{equation}
\begin{equation}
\label{Eq:nb}
-e n_b=\epsilon_0\epsilon_r^{hBN}(E_{b}-E_{gg}),
\end{equation}
where the vacuum permittivity  is given as $\epsilon_0=8.854\times10^{-12}\,F/m$. Further the electric fields are given by the voltage differences between the layers and leads to the following sets of equations:

\begin{equation}
\label{Eq:Eb}
E_b d_b = V_b-V_{bg}
\end{equation}
\begin{equation}
\label{Eq:Egg}
E_{gg} d_{gg} = V_t-V_{b}
\end{equation}
\begin{equation}
\label{Eq:Et}
E_t d_t + E_{Al_2O_3} d_{Al_2O_3} = V_{tg}-V_{t}.
\end{equation}

The magnitude of the electric field between the two graphene sheets follows from Eq.\ref{Eq:Egg} and \ref{Eq:el-chem}:

\begin{equation}
\label{Eq:Egg2}
E_{gg} = \frac{V_t-V_b}{d_{gg}}=\frac{\sqrt{\pi} \hbar v_F}{e d_{gg}}\Big(\sgn(n_t)\sqrt{|n_t|}-\sgn(n_b)\sqrt{|n_b|}\Big).
\end{equation}

From Eq.\ref{Eq:Eb} we obtain that $V_{bg}=V_b-E_b d_b$, while $E_b$ can be expressed as a function of $n_b$ and $E_{gg}$ using Eq.\ref{Eq:nb}. Therefore it follows that $V_{bg}(n_t,n_b)$ is given as,

\begin{equation}
\label{Eq:Vbg1}
\begin{split}
V_{bg} &=V_b+d_b\Big(\frac{en_b}{\epsilon_0\epsilon_r^{hBN}}-E_{gg}\Big) \\ &= \frac{\sgn(n_b)\sqrt{\pi} \hbar v_F}{e}\sqrt{|n_b|}+\frac{e n_b d_b}{\epsilon_0\epsilon_r^{hBN}}-\frac{\sqrt{\pi}\hbar v_F d_b}{e d_{gg}}\Big(\sgn(n_t)\sqrt{|n_t|}-\sgn(n_b)\sqrt{|n_b|}\Big).
\end{split}
\end{equation}

The same can be done for $V_{tg}=V_t+E_t d_t + E_{Al_2O_3} d_{Al_2O_3}$ starting from Eq.\ref{Eq:Et}. By using the relation between the two electric fields one obtains that,

\begin{equation}
\label{Eq:Vtg1}
V_{tg}=V_t+E_t\Big(d_t+\frac{\epsilon_r^{hBN}}{\epsilon_r^{Al_2O_3}}d_{Al_2O_3}\Big)
\end{equation}

For simplification we define $d_t^{eff}=d_t+\frac{\epsilon_r^{hBN}}{\epsilon_r^{Al_2O_3}}d_{Al_2O_3}$. Again we can replace $E_t$ with Eq.\ref{Eq:nt} and in the and we obtain the result,

\begin{equation}
\label{Eq:Vtg}
V_{tg}= \frac{\sgn(n_t)\sqrt{\pi} \hbar v_F}{e}\sqrt{|n_t|}+\frac{e n_t d_t^{eff}}{\epsilon_0\epsilon_r^{hBN}}-\frac{\sqrt{\pi}\hbar v_F d_t^{eff}}{e d_{gg}}\Big(\sgn(n_b)\sqrt{|n_b|}-\sgn(n_t)\sqrt{|n_t|}\Big).
\end{equation}

Eq.\ref{Eq:Vbg1} and \ref{Eq:Vtg} are giving the relation between the gate voltages and the charge carrier densities. To obtain now the charge carrier densities for two given voltages the equations were inverted numerically.

\subsection*{Measurement of $I_c$}
$I_c$ was measured using a  FCA3000 counter. The appearance of a finite voltage ($V_{trig}$) over the junction was triggered, while sweeping the bias current. The value of $V_{trig}$ was set to 6$\mu$V to be able to measure $I_c$ in the entire gate range, e.g. small $I_c$ at the DP, since the smallest detectable value is given by $I_c^{min}=\frac{V_{trig}}{R_n^{DP}}=2.5$\,nA. This small trigger voltage makes the measurement sensitive to voltage noise, which can lead to a trigger error and results in a reduced value for $I_c$. Due to this, the maximum value of hundred measurements of the stochastic switching current is taken, which deviates not more then 10$\%$ from the mean value, since the mean value also contains the trigger errors.

\subsection*{$R_nI_c$ of $J_2$}

The product of $R_n$ and $I_c$ in a JJ in the long regime, namely that the junction length is larger than the superconducting coherence length ($\xi_s$), is proportional to the Thouless energy ($E_{th}$) \cite{Borzenets2016, Dubos2001}. This energy is inversely proportional to the time ($\tau$) that a charge carrier spends in the junction, i.e. the graphene. For a ballistic junction $E_{th}=\hbar v_F/L$, where $L$ the junction length.

\begin{figure}[htb!]
	\centering
	\includegraphics[scale=0.5]{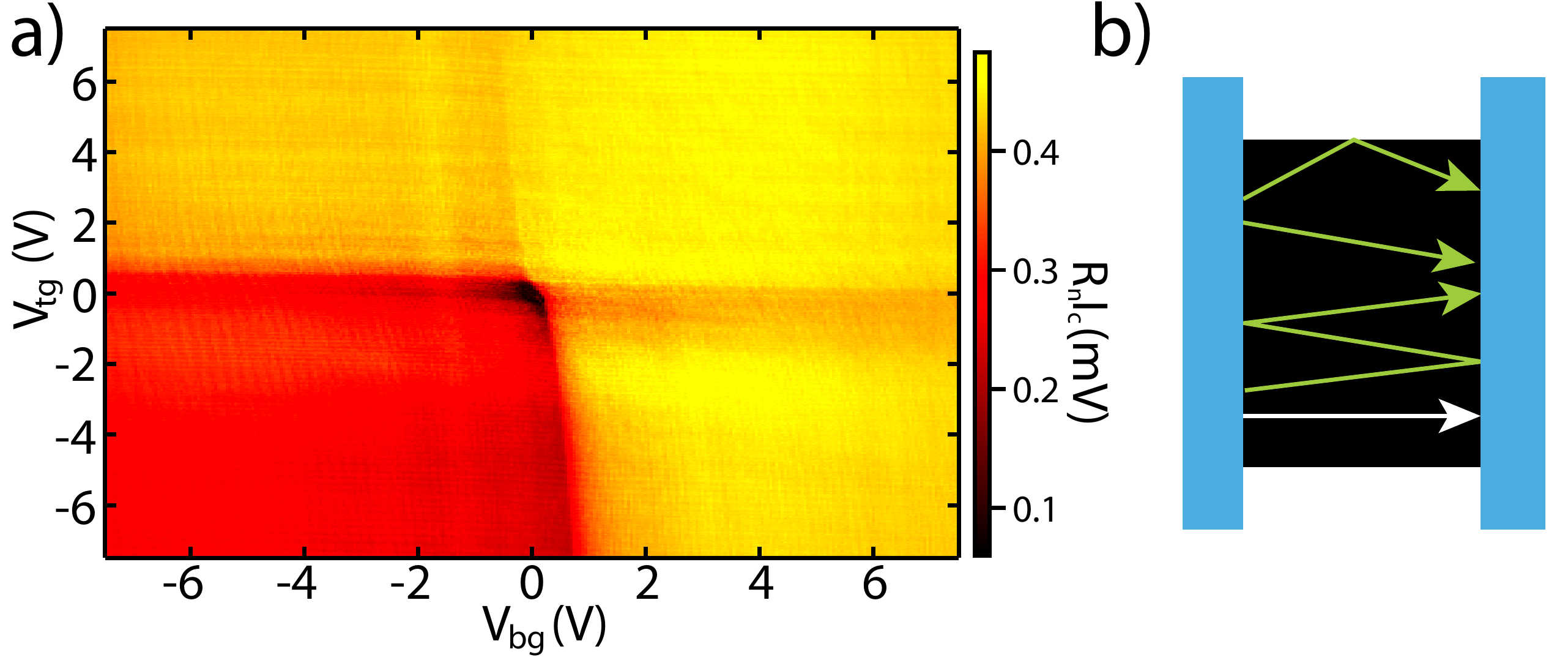}
	\caption{a) Product of $R_n$ and $I_c$ as a function of $V_{tg}$ and $V_{bg}$. If both layers are n doped we observe a value of 0.4\,mV. This value reduces to 0.25\,mV, if both layers are p doped. b) Schematic drawing of a Josephson junction and possible ballistic trajectories of the supercurrent carrying channels. The superconducting leads are indicated by blue, while the graphene is black. The white arrow corresponds to the shortest trajectory between the leads, while  scattering at the physical edge of the graphene, reflection at the imperfect contacts or a finite angle distribution can lead to increased length of the paths.}
	\label{fig:RnIc}
\end{figure} 

$R_nI_c$ for J$_2$ is shown in Fig.\ref{fig:RnIc}\,a. We observe a constant value in each quadrant apart the DPs, i.e. nn, np, pn, and pp, for the product, which indicates a constant value of $E_{th}$. Nevertheless, the value never reaches  the expected one of 1\,mV given by $v_F=10^6$\,m/s for graphene and the junction length $L=650$\,nm. This points into the direction, that the superconducting path decohere stronger than expected. Further, $R_nI_c$ varies between the different quadrants. While it is equal to 0.4\,mV, if both layers are n doped, it decreases to 0.25\,mV in the pp regime. This points into the direction that the imperfect contacts are leading to a suppression of $E_{th}$. This can be seen as the charge carrier are spending an effectively longer time than $L/v_F$ in the junction, which can be due to reflection at the contacts or the physical graphene edges (see Fig.\ref{fig:RnIc}\,b).

\subsection*{Suppressed resistance in moderate out-of-plane magnetic fields}
\begin{figure}[htb]
	\centering
	\includegraphics[scale=1]{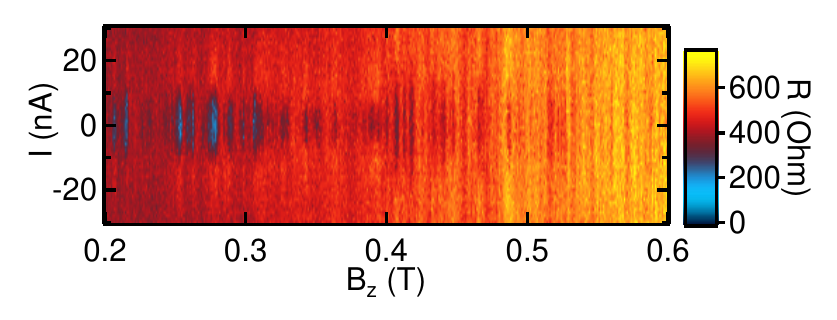}
	\caption{Resistance as a function of current bias and out-of-plane magnetic field. A suppression of the resistance around 0\,nA bias current was observed up to fields of 440\,mT.}
	\label{fig:R_Bz}
\end{figure}

At magnetic fields large enough to suppress the supercurrent in the JJ but smaller than the field needed to be in the quantum Hall (QH) regime, irregular oscillations of the resistance around zero dc-current bias are observed (see Fig.\ref{fig:R_Bz}). The gate voltages were set to $V_{bg}$=-1.5\,V and $V_{tg}$=1.5\,V, while the resistance as a function of out-of-plane magnetic field ($B_z$) was measured with a standard lockin technique. The ac-current amplitude was set to 50\,pA. The appearance of these random oscillations were already observed by Ben Shalom et al.\cite{BenShalom2016} and are attributed to the ballistic transport nature of the junction, which leads to billiard like trajectories at the edges of the sample. These trajectories are suspected to form irregularly Andreev states, while the ones in the bulk are fully suppressed by the magnetic field. Therefore, it is another indication of the ballistic transport nature of J$_2$. Nevertheless, the observation of this superconducting pockets disappears at fields larger than 440\,mT and were not observed in the QH regime within our resolution of 50\,pA as shown in Ref.\cite{Amet2015}.

\subsection*{Calibration and alignment of the in-plane magnetic field}
\begin{figure}[htb]
	\centering
	\includegraphics[scale=1.5]{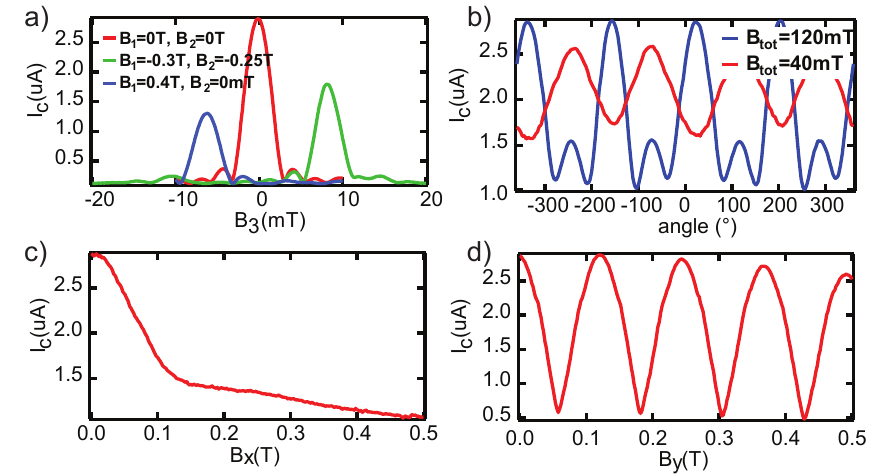}
	\caption{a) Critical current as a function of $B_3$ for three pairs of $B_1$ and $B_2$. b) Critical current as a function of the direction of the in-plane field for two fixed magnitudes of the in-plane magnetic field. c) Dependence of the critical current as a function of $B_x$. d) Critical current as a function of $B_y$.}
	\label{fig:Balign}
\end{figure}

To measure the in-plane magnetic field dependence of such a DLG SQUID device, one has to carefully calibrate and adjust the direction of the magnetic field. Due to the large ratio between the JJs area and the area of the SQUID loop, e.g. 1:35 for J$_2$, the $I^i_c$ of each JJ is more sensitive to an out-of-plane magnetic field than $I_c$ of the SQUID to an in-plane field. If the alignment is imperfect, which results in a finite out-of-plane component, the SQUID interference pattern decays due to the interference of the supercurrent in the individual junctions.  Further we will show that also a component x-direction (see Fig.\ref{fig:Balign}\,c) leads to a reduction in $I_c$ as well.

The calibration was performed using a 3D vector magnet with the magnetic fields $B_1$, $B_2$, and $B_3$, which are perpendicular to each other. While the graphene plane was roughly lying in the plane of the first and second magnet with $B_1$ and $B_2$, the magnetic field of the third one is pointing out-of-plane. In a first step we had to measure three different points, which are in the xy-plane of the sample. This was done by setting the values of the first and second magnet to the values given in the inset of Fig.\ref{fig:Balign}\,a. At each point the out-of-plane magnetic field ($B_3$) was swept and a Fraunhofer like interference pattern was measured. The point in $B_3$, where $I_c$ is maximal reflects the best compensation of the out-of-plane magnetic field, i.e. correspond to a magnetic field in the plane of the JJs. With these three point we defined two vectors, which have to lie in-plane of the graphene layers. To define now a coordinate system we took the cross product of these two vectors to obtain the normal vector $\vec{n}$ of the plane. Then one of the original vectors was normalized and defined as the temporally x-axis ($\vec{e_x}$). By taking now the cross product of $\vec{e_x}$ and $\vec{n}$ we obtain the unit vector in y-direction ($\vec{e_y}$). The two unit vectors $\vec{e_x}$ and $\vec{e_y}$ span now the plane of the graphene layers and allows us to sweep the magnetic field in this plane. Notice, that the direction of the defined vectors are arbitrary and not related to any alignment with the device, e.g. contacts, yet. To find calibrate the magnetic field direction with respect to the device structure, we rotated the magnetic field from -360$^\circ$ to 360$^\circ$ for two different magnitudes (see Fig.\ref{fig:Balign}\,b). Curves with a periodicity of 180$^\circ$ were observed as expected, but the origin of their shape was not fully clear in the beginning. Therefore the magnetic field direction was fixed at an angle of a maxima of either curve shown in Fig.\ref{fig:Balign}\,b. By sweeping the magnitude of the magnetic field in these two direction we observed the interference pattern plotted in Fig.\ref{fig:Balign}\,c and d, from which we could determine the in-plane field direction perpendicular to the SQUID ($B_y$). Notice, that $I_c$ also strongly depends on the magnitude of the magnetic field which is applied parallel to the SQUID's cross section, i.e. in supercurrent direction. This suppression by $B_x$ is attributed to the Meissner effect, which expels the  magnetic field out of the superconducting contact leading to a finite and inhomogeneous out-of-plane magnetic field through the graphene planes. In the direction of $B_y$ we find the modulation of $I_c$ typical for a SQUID. A small decay of the maximal value is observed at higher fields \cite{Steinigeweg2017}. This can either come from a magnetic field component in x or z-direction due to an imperfect alignment or due to out-of-plane corrugations of the individual graphene layers \cite{Couto2014, Kim2019}.

To obtain the CPR, we subtracted the average of $I_c$ over one period for every value of $n_i$ and $B$, which corresponds to the switching current of the reference junction. This leaves us with the CPR.

\subsection*{Minima of $I_c(B_y)$ as a function of $V_{tg}$}
For a symmetric SQUID ($I^1_c=I^2_c$) with a sinusoidal CPR, one expects a $|\cos(\pi \Phi/\Phi_0)|$ like interference pattern of $I_c$ vs magnetic field. Therefore one would observe that $I_c$ fully vanishes at a magnetic flux equal to $\Phi_0$/2. If the SQUID is not symmetric, the supercurrent flowing in the two JJs will not fully compensate each other at $\Phi_0$/2, leaving us with a finite $I_c$. But this observation is also possible if the CPR is not sinusoidal, even if the JJs are symmetric. To show that the non vanishing $I_c$ in the interference pattern in Fig.3\,d of the article, is not only due to an asymmetry of the JJ, but rather given by a non sinusoidal CPR, we measured $I_c$ as a function of $V_{tg}$, while $V_{bg}$ was fixed at 5\,V and the in-plane magnetic field at -181.6\,mT, which corresponds to a minimum of the interference pattern (see Fig.\ref{fig:Balign}). When $V_{tg}$ is tuned mainly the critical current carried by the top graphene layer changes. Like this it is possible to change between a symmetric and an asymmetric SQUID configuration. At the DP of the top layer ($V_{tg}\approx 0$\,V) the supercurrent is carried only by the bottom layer and the critical current is therefore only given by $I^b_c$. When the gate voltage is increased, $I^t_c$ also increases. Since there is a phase difference of roughly $\pi$/2 between the JJs due to the magnetic flux, the supercurrent flows in the opposite direction, which leads to a decrease of SQUID's $I_c$. This trend continues until $I^t_c=I^b_c$, where $I_c$ will reach its minimum before it starts to increase again due to opposite asymmetry ($I^t_c>I^b_c$). The non zero $I_c$ in the symmetric SQUID is attributed to the non sinusoidal CPR observed and discussed in the main text. This can be seen by taking a look how $I_c$ as a function of $B$ is calculated. First, the total supercurrent is given by $I_c=I^t_c f^t(\varphi_t)+I^b_c f^b(\varphi_b)$, where $\varphi_t$ ($\varphi_b$) is the phase difference over the top (bottom) JJ and $\varphi_t=\varphi_b+\pi \Phi/\Phi_0$. For a given magnetic field $\varphi_t-\varphi_b$ is fixed but not the value of $\varphi_b$. To obtain now $I_c$ one has to maximize $I_c$ over $\varphi_b$. Therefore, to obtain a $I_c$ of zero, $I_c$ has to be zero for all $\varphi_b$. This is the case for the sum of two sinus curves shifted by $\pi$/2 but is never the case if the CPRs are skewed sinusoidal functions. Therefore, the non vanishing supercurrent can be attributed to a non sinusoidal CPR.

\begin{figure}[htb]
	\centering
	\includegraphics[scale=1]{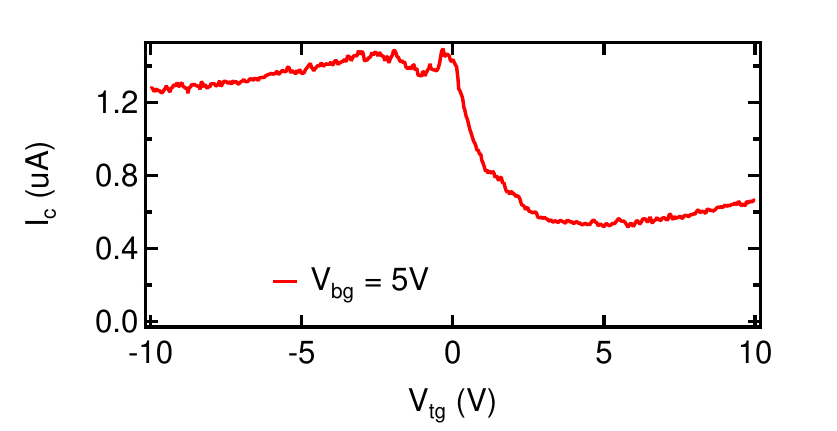}
	\caption{$I_c$ as a function of $V_{tg}$ for $V_{bg}$=5\,V. The critical current was measured for a in-plane magnetic field ($B_y$) of -181.6\,mT.}
	\label{fig:minIc}
\end{figure}

\section*{Gate dependence of $a_n$}
\begin{figure}[htb]
	\centering
	\includegraphics[scale=1.5]{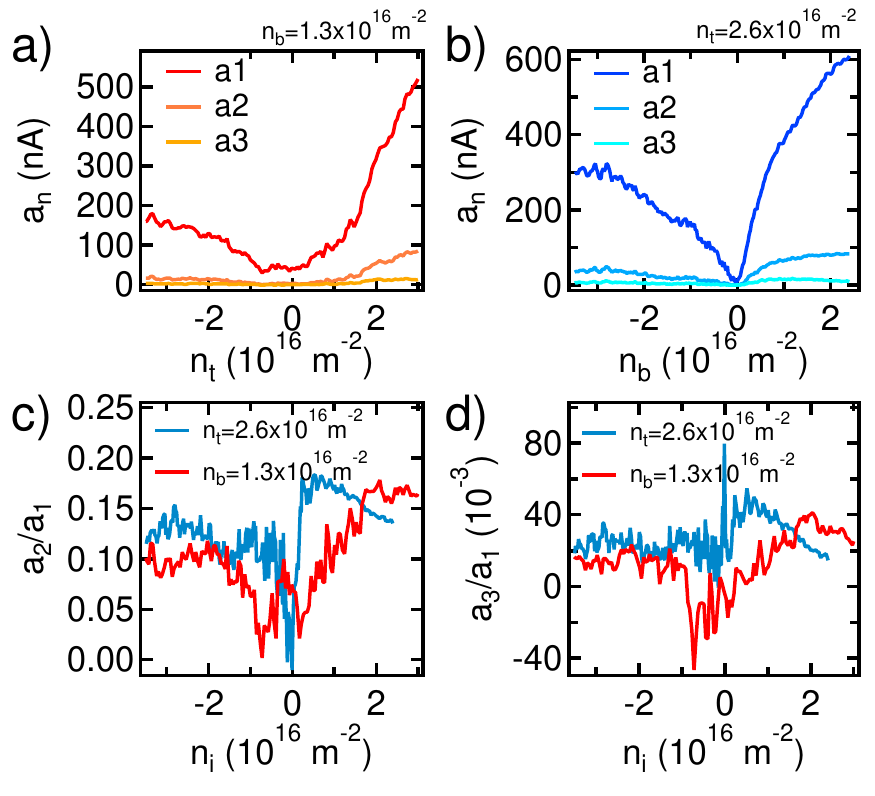}
	\caption{a) Fitting coefficients $a_1$, $a_2$, and $a_3$ as a function of $n_t$ for $n_b$=1.3$\times$10$^{-2}$. b) Fitting coefficients $a_1$, $a_2$, and $a_3$ as a function of $n_b$ for $n_t$=2.6$\times$10$^{-2}$. c) Ratio of $a_2$ and $a_1$, which are shown in a) and b). d) Ratio of $a_3$ and $a_1$, which are shown in a) and b).}
	\label{fig:ratios}
\end{figure}

To fit the CPR we used Eq.3 in the article, which contains up to the fifth harmonic in frequency. If the CPR is sinusoidal $a_2$ to $a_5$ are all  zero and only the first harmonic exists. The non-vanishing of the higher harmonic amplitudes indicates, that the CPR will be skewed and can be used as an alternative measurement quantity to the skewness (see main text) to define the deviation of the CPR from the sinusoidal behavior. For completion, we plot the $a_1$ to $a_3$ and the ratio between $a_2$ and $a_1$, as well as the ratio between $a_3$ and $a_1$ in Fig.\ref{fig:ratios}. The amplitudes $a_4$ and $a_5$ are much smaller then the the others and their contribution to skewness of the CPR can be  neglected. For n (p) doped graphene a skewness of 0.25 (0.15) was extracted. This value corresponds to a ratio of $a_2$/$a_1$ of around 0.15 (0.1).

\subsection*{Calculation of the interference pattern}
\begin{figure}[htb]
	\centering
	\includegraphics[scale=0.8]{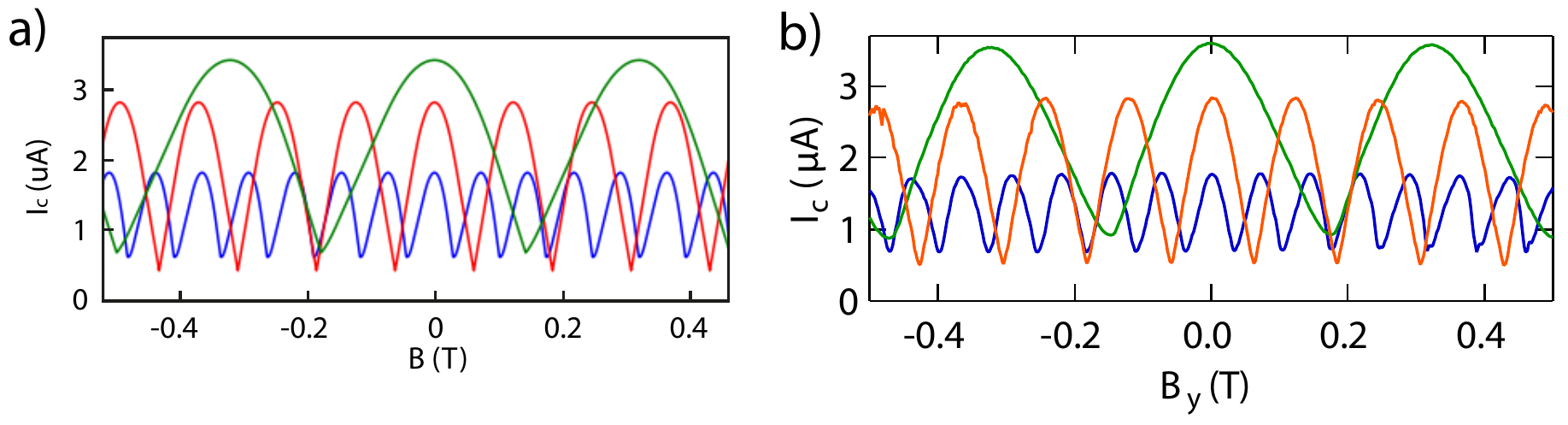}
	\caption{a) Calculated interference pattern for a SQUID with a skewed CPR ($S$=0.18). The different curves, indicated by different colors, were obtained for different SQUID areas, which corresponds to the ones of J$_1$ (blue), J$_2$ (red) and J$_3$ (green). b) Measurement of the critical current as a function of in-plane magnetic field for J$_1$, J$_2$, and J$_3$.}
	\label{fig:CalIn}
\end{figure}

To get an idea of the asymmetry of the measurement shown in Fig.3\,d of the main article, we calculated the in-plane magnetic field dependence of the $I_c$. The CPRs were chosen to be equal and with skewness of $S$=0.18, which is given by the choice of the prefactors $a_1$, $a_2$, and $a_3$ in Eq.\,3 of the main text. The blue curve is the result of $I^b_c=1.2\,\mu$A and $I^t_c=0.6\,\mu$A, the red one for  $I^b_c=1.4\,\mu$A and $I^t_c=1.4\,\mu$A and the green for  $I^b_c=2\,\mu$A and $I^t_c=1.4\,\mu$A. For the blue result we took the dimension (junction length and middle hBN thickness) of J$_1$, for the red the dimension of J$_2$, and for the green curve the dimension of J$_3$. The result reproduces qualitatively the measurements in Fig.\ref{fig:CalIn}\,b. Therefore, we conclude that the in-plane magnetic field dependence of J$_2$ was in a rather symmetric state of the SQUID, while for J$_1$ and J$_3$ the SQUID was slightly asymmetric. The calculations also reproduce the shape of the different curves.

\subsection*{Estimation of the loop inductance}
The loop inductance ($L_s$) can lead to  screening of the external magnetic field, which modifies the actual flux ($\Phi$) inside the SQUID. Further it makes the relation between $\Phi$ and the external flux ($\Phi_{ext}$) non linear. When the magnetic field axis is converted to a phase axis, this non linearity has to be taken into account, if $L_s$ or $I_c$ is large. In the case of a symmetric dc-SQUID, $\Phi_{ext}$ as a function of $\Phi$  can be expressed by

\begin{equation}
\label{eq:phi_ext}
\Phi_{ext}=\Phi+L_sI_c f\left(\frac{\pi \Phi}{\Phi_0}\right).
\end{equation}

Further, $L_s$ and $I_c$ defining the limit, at which the phase biasing by a magnetic field gets hysteretic. This limit is given by $\pi L_sI_c/\Phi_0\approx 1$. To estimate the loop inductance, we calculated the kinetic inductance ($L_k$) of the MoRe leads and the geometric inductance ($L_g$) of the SQUID loop. The sum of these inductances results in $L_s$.

$L_k$ was measured by the temperature dependence of the resonance frequency ($f_{res}$) of a $\lambda$/4-resonator.

\begin{equation}
\label{eq:fres}
f_{res}=\frac{1}{4l\sqrt{\left(L_m+\frac{L^0_k}{1-\left(\frac{T}{T_c}\right)^4}\right)\cdot{}C_m}},
\end{equation}

where $l$ is the length of the resonator and $L_k^0$ is the kinetic inductance per unit length in the zero temperature limit. The geometric inductance of the resonator ($L_m$) as well as the geometric capacitance of the resonator ($C_m$)  were calculated as described in Ref.\cite{Gevorgian1994}. We obtain a sheet inductance of $L_{k}^s$=4.26\,pH for a resonator thickness of 70\,nm. $L_k$ is obtained by multiplying the sheet inductance with the interlayer distance, here $d_{gg}$=25\,nm, and divide it by the the contacts width of 550\,nm. By doing so $L_k$=0.19\,pH. Note, that this is an upper bound of the kinetic inductance, since the $L_k^s$ was determined using a 70\,nm thick resonator. Here, due the supercurrent direction the thickness would correspond to the length of the contact region, which is about 850\,nm. Therefore, we expect the kinetic inductance to be even smaller.

To estimate $L_g$ we calculate the inductance of a rectangular loop as derived in Ref.\cite{Shatz2014}. Here we take the following values: $l_1$=$L$, $l_2=d_{gg}$, $w$=0.3\,nm (thickness of graphene) and $h$=1\,nm the width of the of the loop. By taking $h$ equal to only 1\,nm instead of the entire junction width, we get an upper limit of the geometrical inductance of $L_g$=1.2$\times 10^{-12}$\,H. This has to be done since the used formula does not hold if $h$ is much larger then the product of $l_1$ and $l_2$.

We calculate now the difference between $\Phi_{ext}$ and $\Phi$ at $\Phi=\Phi_0/2$, where the effect of the screening is the strongest. For a critical current of 3\,$\mu$A, the difference is not more then 0.3$\%$. Further, the maximal current, which can be passed through the SQUID before it starts to behave hysteretic is $I_c^h\approx 0.5$\,mA. For these reasons screening effects can be neglected in our measurements, since the measured critical currents are way smaller and the non linearity is not present.

\bibliography{refs}